\documentclass[12pt,a4paper]{article}

\usepackage{amsmath}
\usepackage{latexsym}

\def\Q{{\bf Q}}
\def\K{{\bf K}}
\def\S{{\hat S}}
\def\V{{\hat V}}
\def\T{{\hat T}}
\def\a{{\alpha}}
\def\b{{\beta}}

\def\ord{{\rm ord}}
\def\rg{{\it rGCD}}
\def\lg{{\it lGCD}}
\def\ll{{\it lLCM}}
\def\rl{{\it rLCM}}
\def\t#1#2#3{#1\stackrel{#2}{\longrightarrow}{#3}}

\newtheorem{theorem}{Theorem}

\title{Factorization in categories of systems of linear partial differential equations}
\author{Sergey P. Tsarev\thanks{This paper was written with partial financial
support from the RFBR grant 06-01-00814.}\\
Krasnoyarsk State Pedagogical University\\
Lebedevoi, 89 \\
 Krasnoyarsk, 660049, \\
 Russia \\
 {\tt sptsarev@mail.ru} }

\date{26 December 2007}

\begin{document}

\maketitle

%---------------------------------------------------------------------

\section{Introduction}

Factorization of systems of differential equations was first studied
for the case of a single linear ordinary differential equation
(LODE) with linear ordinary differential operator (LODO) of the form
\begin{equation}\label{op1}
  L=f_0(x) D^n+f_1(x)D^{n-1}+\ldots +f_n(x), \quad D=d/dx,
\end{equation}
where the coefficients $f_s(x)$ belong  to some differential field
$\K$. Factorization is a useful tool for computing a closed form
solution of the corresponding  linear ordinary differential equation
$Ly =0$ as well as determining its Galois group (see for example
\cite{S93,SU,vPS}).
%%%%%%
For simplicity and without loss of generality we suppose that
operators (\ref{op1}) are reduced (i.e. $f_0(x) \equiv 1$) unless we
explicitly state the reverse. The most popular case of the
differential field $\K=\bar \Q(x)$ of rational functions with
rational or algebraic number coefficients is a nontrivial example
which is well investigated and will be considered hereafter when we
discuss any constructive results.

In this paper we give a review of the current state of the theory of
factorization of ordinary and partial differential operators and
even more generally, of systems of linear differential equations of
arbitrary type (determined as well as overdetermined). We start with
elementary algebraic theory of factorization of linear ordinary
differential operators (\ref{op1}) developed in the period
1880--1930. After exposing these classical results we sketch more
sophisticated algorithmic approaches developed in the last 20 years.
The revival of this theory in the last two decades is motivated by
the development of powerful computer algebra systems and
implementation of nontrivial algebraic and differential algorithms
such as factorization of polynomials and indefinite integration of
elementary functions.

 The main part of this paper will be devoted to
modern generalizations of the  factorization theory to the most
general case of systems of linear partial differential equations and
their relation with explicit solvability of nonlinear partial
differential equations based on some constructions from the ring
theory, theory of partially ordered sets (lattices) and that of
abelian categories. Many of the results of this paper may be exposed
within the framework of the Picard-Vessiot theory. But we follow a
much simpler algebraic approach in order to facilitate the
aforementioned generalizations.

The proper theoretical background for the simplest
case---factorization of linear ordinary differential operators with
rational coefficients---was known already in the end of the XIX
century. Paradoxically,  mathematicians of that epoch had developed
even a nontrivial (theoretical) algorithm of factorization of such
operators \cite{Beke}! A review of this theory can be found in
\cite{schl}. In contrast to the well-known property of uniqueness of
factorization of usual commutative polynomials into irreducible
factors,  a simple example $D^2 = D\cdot  D = (D+ 1/(x-c))\cdot
(D-1/(x-c))$ shows that some LODO may have essentially different
factorizations with factors depending on some arbitrary parameters.
Fortunately according to the results by E.~Landau \cite{Lan} and
A.~Loewy \cite{Loewy1,Loewy2} exposed below all possible
factorizations of a given operator $L$ over a fixed differential
field have the same number of factors in different expansions $L=
L_1\cdot  \cdots \cdot L_k = \overline L_1\cdot  \cdots \cdot
 \overline L_r $ into irreducible factors and the factors $L_s$,
$\overline L_p$ are pairwise "similar". (Hereafter we always suppose
 the order of factors to be
greater than $0$: $\ord(L_i)>0$, $\ord(\overline L_j)>0$). We
outline the main ideas of this classical theory in
Section~\ref{sec1}. For simplicity we discuss here only the case of
differential operators, a generalization for the case of a general
Ore ring (including difference and $q$-difference operators, see
\cite{BP94,BP}) is straightforward.

Subsequent Sections are devoted to different aspects of the theory
of factorization of linear partial differential operators.

\section{Factorization of LODO}\label{sec1}

The basics of the algebraic theory of factorization of LODO was
essentially given already in \cite{Loewy1,Loewy2},
\cite{Ore1}--\cite{Ore4}. Algebraically the main results are just an
easy consequence of the fact that the ring $\K[D]$ of LODO with
coefficients in a given differential field $\K$ is Euclidean: for
any LODO $L$, $M$ there exist unique LODO $Q$, $R$, $Q_1$, $R_1$
such that
$$L=Q\cdot M + R, \  L=M\cdot Q_1 + R_1, \  ord(R)<ord(M),
 \  ord(R_1)<ord(M).$$
For any two LODO $L$ and $M$ using the right or left Euclidean
algorithm one can determine their right greatest common divisor
$\rg(L,M)=G$, i.e. $L=L_1\cdot  G$, $M= M_1\cdot  G$ (the order of
$G$ is maximal) and their right least common multiple $\rl(L,M)=K$,
i.e. $K=\overline M \cdot  L= \overline L \cdot  M$ (the order of
$K$ is minimal) as well as their left analogues \lg\ and \ll. All
left and right ideals of this ring are principal and all two-sided
ideals are trivial. Operator equations
 \begin{equation}
 X\cdot  L + Y\cdot  M = B, \qquad L\cdot  Z + M\cdot  T = C
% No 2
\end{equation}
 with unknown operators $X$, $Y$, $Z$, $T$ are solvable iff
$\rg(L,M)$ divides $B$ on the right and $\lg(L,M)$ divides $C$ on
the left. We say that an operator $L$ is (right) {\it transformed
into $L_1$ by an operator ({\it not necessary reduced})} $B$, and
write $\t LB{L_1}$, if $\rg(L,B)=1$ and $K= \rl(L,B) = L_1\cdot  B =
B_1 \cdot L$. In this case any solution of $Ly=0$ is mapped by $B$
into a solution $By$ of $L_1y=0$. Using (2) one may find with
rational algebraic operations an operator $B_1$ such that
$\t{L_1}{B_1}L$, $B_1\cdot B = 1 ({\rm mod} L)$. Operators $L$,
$L_1$ will be also called {\it similar} or {\it of the same kind}
(in the given differential field \K). So for similar operators the
problem of solution of the corresponding LODE  $Ly = 0$, $L_1y =0$
are equivalent. One can define also the notion of left-hand
transformation of $L$ by $B$ into $L_1$: $K= \ll(L,B) = B\cdot  L_1
= L\cdot  B_1$. Obviously left- and right-hand transformations are
connected via the adjoint operation. Also one may prove
(\cite{Ore3}) that two operators are left-hand similar iff they are
right-hand similar. A (reduced) LODO is called {\it prime} or {\it
irreducible} (in the given differential field \K) if it has no
nontrivial factors aside from itself and $1$. Every LODO similar to
a prime LODO is also prime. Two (prime for simplicity) LODO $P$ and
$Q$ are called {\it interchangeable in the product $P\cdot  Q$} and
this product will be called {\it interchangeable} as well if $P\cdot
Q = Q_1\cdot P_1$, $Q_1 \neq P$, $P_1 \neq Q$. In this case $P$ is
similar to $P_1$, $Q$ is similar to $Q_1$ and $\t{P_1}QP$.

\begin{theorem}[Landau \cite{Lan} and Loewy \cite{Loewy1}]
%\label{T1}
Any two different decompositions of a given LODO $L$ into products
of prime LODO $L= P_1 \cdot \cdots \cdot  P_k = \overline P_1 \cdot
\cdots \cdot \overline P_p$ have the same number of factors ($k=p$)
and the factors are similar in pairs (in some transposed order). One
decomposition may be obtained from the other through a finite
sequence of interchanges of contiguous factors (in the pairs
$P_i\cdot P_{i+1}$).
\end{theorem}

All definitions here are constructive over the differential field of
rational functions $\K=\Q(x)$: either using the Euclidean algorithm
or finding rational solutions of LODO one can determine for example
if two given LODO are similar or find all possible (parametric)
factorizations of a given LODO with rational functional coefficients
\cite{ABP,BP94,Bro1,ts96}.

Landau-Loewy theorem also has a useful for the following
\emph{ring-theoretic interpretation}. Namely, every $L \in \K[D]$
generates the corresponding left ideal $|L\rangle$; $L_1$ divides
$L$ on the right iff $|L\rangle \subset |L_1\rangle$. If we have a
factorization $L = L_1 \cdots  L_k$ then we have a chain of
ascending left principal ideals $|L\rangle \subset | L_2 \cdots
L_k\rangle \subset | L_3 \cdots L_k\rangle \subset \ldots \subset|
L_k\rangle \subset| 1\rangle = \K[D]$. If the factors $L_k$ are
irreducible, the chain is maximal, i.e.\ it is not possible to
insert some intermediate ideals between its two adjacent elements.
The Landau-Loewy theorem is nothing but the
\textbf{Jordan-H\"{o}lder -Dedekind chain condition}:
\begin{theorem}\label{JH}
 Any two finite maximal ascending
chains of left principal ideals in the ring $\K[D]$ of LODO have
equal length.
\end{theorem}
Similarity of irreducible factors can be also interpreted in this
approach. Even more general lattice-theoretic interpretation turned
out to be fruitful for a generalization of this simple algebraic
theory for the case of factorizations of \emph{partial} differential
operators \cite{ts98}. Namely, let us consider the set of (left)
ideals in $\K[D]$ as a  partially ordered by inclusion set $\cal M$
(called a poset). This poset has the following two fundamental
properties:

\emph{Property I}) for any two elements $A,B \in \cal M$ (left
ideals!) one can find a unique $C=\sup(A,B)$, i.e.\ such $C$ that $C
\geq A$, $C \geq B$, and $C$ is ``minimal possible''.
 Analogously there exist
a unique $D=\inf(A,B)$,
 $D \leq A$, $D \leq B$, $D$ is ``maximal possible''.

Such posets are called {\em lattices} \cite{grat}. $\sup(A,B)$ and
$\inf(A,B)$ correspond to the GCD and the LCM in $\K[D]$.

  For simplicity (and following the established tradition)
$\sup(A,B)$ will be hereafter denoted as $A+B$ and $\inf(A,B)$ as
$A\cdot B$;

\emph{Property II}) For any three $A,B,C \in \cal M$ the following
{\em modular identity} holds:
$$
(A\cdot C + B)\cdot C = A\cdot C + B\cdot C
$$

Such lattices are called \emph{modular lattices} or \emph{Dedekind structures}.

As on can prove, modularity implies the Jordan-H\"older-Dedekind
chain condition: any two finite maximal chains $ L  > L_1 > \cdots >
L_k > 0 $ and $ L  > M_1 > \cdots > M_r > 0$ for a given $L \in \cal
M$ have equal lengths: $k=r$ (the same for ascending chains). For
the interpretation of the notions of similarity, direct sums, Kurosh
\& Ore theorems on direct sums cf.\ \cite{ts98}.

But even more fruitful for generalizations is the followings
\emph{categorical interpretation} of similarity of LODO and the
Jordan-H\"older-Dedekind chain condition. Namely, let us consider
the following \textbf{abelian category ${\cal LODO}$ of LODO}.

Objects of  ${\cal LODO}$ are \emph{reduced} operators
$L=D^n+a_1(x)D^{n-1}+\ldots +a_n(x),$ $a_i \in \K$. One may ideally
think of  (finite-dimensional!) the solution spaces $Sol(L)$ of such
operators in some sufficiently large Picard-Vessiot extension of the
coefficient field as another way of representation of an object in
${\cal LODO}$. This helps to understand the meaning of some
definitions below, but one shall remember that these solution spaces
are \emph{not constructive} unlike the objects-operators.

Morphisms $Hom(L,L_1)$ in this category are constructively defined
as a \emph{not necessary reduced}  LODO $B$ such that $L_1 \cdot B =
C\cdot L$ for some other LODO $C$. Non-constructively this $B$ may
be seen as a mapping of solutions of $L$ into solutions of $L_1$.
Note that  all operators here have coefficients in some \emph{fixed}
differential field $\K$. Two operators $B_1$, $B_2$ generate the
same morphism iff  $B_1=B_2 (mod\ L)$. Also we should remark that
this definition is \emph{not} equivalent to the definition of a
transformation of operators $\t LB{L_1}$ introduced earlier, because
for morphisms:

1) $B$ and $L$ may have common solutions, i.e.\ a nontrivial
$\rg(B,L)$. This means that the mapping of the solution space
$Sol(L)$ by $B$ may have a kernel $Sol(\rg(B,L))$. The morphism is
not injective in this case.

2) The image of the solution space $Sol(L)$ may be smaller than
$Sol(L_1)$. The morphism is not surjective in this case.

Algebraically this means that $L_1 \cdot B = C\cdot L \neq
\rl(B,P)$.

Similarity of operators $L$ and $L_1$ now simply means isomorphism
of the objects $L$ and $L_1$ in this category.

The following fact is a direct corollary of our representation of
this category as a subcategory of the category of finite-dimensional
vector spaces and linear mappings preserving direct sums, products
etc.:
\begin{theorem}
The category  ${\cal LODO}$  is \emph{abelian}.
\end{theorem}
Among the many useful results for abelian categories (cf.\ for
example \cite{Freyd,GM}) we need the following
\begin{theorem}\label{JH-abel}
Any abelian category with finite ascending chains satisfies the
Jordan-H\"older property.
\end{theorem}
This will serve us as a maximal theoretical framework for an
algebraic interpretation and generalization of the Landau-Loewy
theorem for the case of systems of linear partial differential
equations below. Again there exist notions of direct sums,
Kurosh-Ore theorems on direct sums and the powerful technique of
modern homological algebra for abelian categories \cite{Freyd,GM}.
We will see below that this rather high level of abstraction allows
a very natural generalization of the definition of factorization for
arbitrary systems of linear partial differential equations (LPDE).

%\section{Classical algorithm of factorization (Beke)}
%
%\ {Classical algorithm of factorization (Beke)
%\\ and its modern rivals}
%
%Ideas:
%
%
%
%
%
%\begin{enumerate}
%    \item[(I)]
%$L=L_1\cdot (D-u(x))\Longleftrightarrow u=\frac{y'}{y}$,
%$y(x)$ is a solution of $Ly=0$.
%
%For $u \in k=Q(x)$ this means $y=\exp(\int u \, dx)$
% \\ --- a hyperexponential solution of $Ly=0$.
%
%
%
%    \item[(II)]
% $L=L_1\cdot L_2$, $ord(L_2)=m$
%\\ $\Longrightarrow$ for some \emph{associated} LODO
%$L_{(m)}$,
%\\
%$L_{(m)}=\overline  L_1\cdot (D- \overline  u(x))$
%
%\end{enumerate}
%
%
%

\section{Factorization of LPDO}\label{sec2}

In contrast to the case of ordinary operators, two main results are
seemingly lost for LPDO: Landau-Loewy theorem and the possibility to
use some known solution for factorization of operators. In the case
of a LODO $L$ obviously if one has its solution $L\phi=0$ then one
can split off a first-order right factor: $L=M\cdot
\left(D-\frac{\phi'}{\phi}\right)$. For a LPDO, even if one knows
the \emph{complete} set of solutions, the operator may not be
factorizable, as the following classical examples shows:

\textit{Example 1.} The equation $Lu=\left(D_xD_y
 - \frac{2}{(x+y)^2} \right) u  =0$ with $D_x=\partial/\partial x$, $D_y=\partial/\partial y$
 has the following complete solution:
 $$
Lu=0 \Leftrightarrow  u= -\frac{2 (F(x)+G(y))}{x+y}+F'(x)+G'(y),
$$
where $F(x)$ and $G(y)$ are two arbitrary functions of one variable
each. On the other hand, as an easy calculation shows, the operator
$L$ can not be represented as a product of two first-order operators
(over \emph{any} differential extension of the given coefficient
field $\K=\Q(x,y)$).

\textit{Example 2.} The equation $Lu=\left(D_x D_y
 - \frac{6}{(x+y)^2} \right) u  =0$ again has the following complete solution:
\begin{equation}\label{compsol}
 u= \frac{12 (F(x)+G(y))}{(x+y)^2}-\frac{6 (F'(x)+G'(y))}{x+y}
+F''(x)+G''(y),
\end{equation}
but the operator is again ``naively irreducible''. More generally,
the equation
\begin{equation}\label{cnn+1}
Lu=u_{xy} - \frac{c}{(x+y)^2}u=0, \qquad c=\textrm{const}
\end{equation}
has a complete solution in an ``explicit'' form similar to
(\ref{compsol}) iff $ c=n(n+1)$ for $n \in {\bf N}$. In this case it
has the complete solution in the form
\begin{equation}\label{complSol}
       u = c_0F + c_1F' + \ldots +  c_nF^{(n)}+ d_0 G + d_1 G' +
   \ldots +  d_{n+1} G^{(n+1)}
\end{equation}
with some definite $c_i( x,y)$, $d_i(x, y)$ and  two arbitrary
functions $F( x)$, $ G( y)$.

Only for $c=0$ the corresponding operator is ``naively reducible'':
$L= D_x\cdot D_y $.

The solution technology used here is very old and can be found in
\cite{dar-lec,gour-l} under the name of \emph{Laplace
transformations} or \emph{Laplace cascade method}: after a series of
transformations of (\ref{cnn+1}) one  gets a \emph{naively
factorizable} LPDE!

Another unpleasant example is also ascribed in \cite{blum} to
E.~Landau:  if
\begin{equation}\label{land}
\begin{array}{c}
 P=D_x+xD_y, \quad Q=D_x+1, \\
         R=D_x^2+xD_xD_y+D_x+(2+x)D_y,
\end{array}
\end{equation}
then  $L = Q \circ Q \circ P = R \circ Q$. On the
other hand the operator $R$ is absolutely irreducible, i.e.\ one can
not factor it into product of first-order operators with
coefficients in {\em any} extension of $\Q(x,y)$. So there seems to
be no hope for an analogue of Landau-Loewy theorem for decomposition
of LPDO into product of lower-order LPDO.

We see that a ``naive'' definition of a factorization of a LPDO as
its representation as a product (composition) of lower-order LPDOs
lacks some fundamental properties established in the previous
Section for factorization of LODO.

Recently \cite{ts98} an attempt to give a ``good'' definition of
generalized factorization was undertaken. In the next subsection we
only briefly sketch the ideas of this approach and describe its
nontrivial relation to explicit integrability of \emph{nonlinear}
partial differential equations.

\subsection{General theory of factorization of an arbitrary single
LPDO, ring-theoretic approach}

Our goal in \cite{ts98} was  to define a notion of factorization
with ``good'' properties:
\begin{itemize}
\item Every LPDO $L$ shall have only finite chains of ascending
generalized factors.  In particular $D_x$ should be irreducible.

\item  Jordan-H\"{o}lder property: all possible generalized
factorizations of a given  operator $L$ have the same number of
``factors'' in different expansions into irreducible factors and the
``factors'' should be pairwise ``similar'' in such expansions.

\item Existence of \emph{large} classes of solutions should be related to
factorization.

\item Classical theory of integration of LPDO using the Laplace cascade
method should be an integral part of this generalized definition.
\end{itemize}
An obvious extension of the definition may be suggested if one will
use ascending chains of arbitrary (not necessary principal) left
ideals starting from the left ideal generated by the given operator:
\begin{equation}\label{chainsI}
|L\rangle \subset I_1 \subset I_2 \subset \ldots \subset I_k \subset
|1\rangle.
\end{equation}
Unfortunately  one can easily see that even for the operator $D_x$
we have such chains, and they have unlimited length: $|D_x\rangle
\subset |D_x,D_y^m\rangle \subset|D_x,D_y^{m-1}\rangle \subset
\ldots |D_x,D_y\rangle \subset |1\rangle$! So we shall take some
special class of ideals, more general, than the principal ideals,
but much less rich then arbitrary left ideals. In \cite{ts98} we
gave a definition of such a suitable subclass of left ideals called
\emph{divisor ideals}.

For such special left ideals of the ring of LPDO:
\begin{itemize}
\item chains (\ref{chainsI}) will be finite and different maximal chains
for a given $L$ they have the same length: if $|L\rangle \subset I_1
\subset I_2 \subset \ldots \subset I_k \subset |1\rangle $,
$|L\rangle \subset J_1 \subset J_2 \subset \ldots \subset J_m
\subset |1\rangle $, then $ k=m$ and one can prove a natural
lattice-theoretic ``similarity'' of ``factors'' in both chain.

\item  Irreducible LODO will be still irreducible as LPDO.

\item For $dim=2$, $ord=2$ (that is for operators with two independent
variables of order two) a LODO is factorizable in this generalized
sense (i.e.\ having a nontrivial chain (\ref{chainsI})) iff it is
integrable with the Laplace cascade method. We describe this cascade
method below in subsection~\ref{ssecLaplace}

\item Algebraically, the problem is reduced from the ring $Q(x,y)[D_x,D_y]$ to
factorization in rings of formal LODO with noncommutative
coefficients $Q(x,y,D_x)[D_y]$ and/or $Q(x,y,D_y)[D_x]$ (Ore
quotients); in these rings all left and right ideals are again
principal ideals.
\end{itemize}
The details, rather involved, may be found in \cite{ts98}. This
approach nevertheless suffers from the following problems:
\begin{itemize}
\item   The definition of divisor ideals given in \cite{ts98} is
very technical, not intuitive.

\item  No algorithms for such generalized factorization is known.

\end{itemize}
A generalization of this \emph{ring-theoretic approach}  to
\emph{systems} of LPDE was proposed recently by M.~Singer. Another
ring-theoretic approach was considered in \cite{GSch05}.

In the next subsection we propose a different, much more intuitive
definition of generalized factorization.

%---------------------------------------------------------------------

\subsection{General theory of factorization of  arbitrary systems of
LPDE, approach of abelian categories}

Abelian category ${\cal SLPDE}$ of arbitrary  systems of LPDE is
defined by its objects which are simply systems
 \begin{equation}\label{sysLPDE}
 S: \left\{\begin{array}{l}
L_{11}u_1 + \ldots + L_{1s}u_s=0,\\
\cdots \\
L_{p1}u_1 + \ldots + L_{ps}u_s=0,\\
\end{array}
\right.% \quad
\begin{array}{l}
L_{ij} \in Q(x_1,\ldots,x_n)[D_{x_1}, \ldots, D_{x_n}], \\
u_k=u_k(x_1,\ldots,x_n).
\end{array}
 \end{equation}
Morphism $P: S \rightarrow Q$ of two systems is defined as a matrix
of differential operators
\begin{equation}\label{morP}
P:\left\{\begin{array}{l}
v_1=P_{11}u_1 + \ldots + P_{1s}u_s,\\
\cdots \\
v_m=P_{m1}u_1 + \ldots + P_{ms}u_s,\\
\end{array}
\right.
\end{equation}
$P_{ij} \in Q(x_1,\ldots,x_n)[D_{x_1}, \ldots, D_{x_n}]$ with the
condition that any solution set $\{u_1, \ldots, u_s\}$ of the source
system (\ref{sysLPDE}) is mapped into a subspace of the solution
space $\{v_1, \ldots, u_m\}$ of the target system
 \begin{equation}\label{sysLPDEQ}
 Q: \left\{\begin{array}{l}
M_{11}u_1 + \ldots + M_{1m}v_m=0,\\
\cdots \\
M_{q1}v_1 + \ldots + M_{qm}v_m=0,\\
\end{array}
\right.% \quad
\begin{array}{l}
M_{ij} \in Q(x_1,\ldots,x_n)[D_{x_1}, \ldots, D_{x_n}], \\
v_k=v_k(x_1,\ldots,x_n).
\end{array}
\end{equation}
The standard differential Groebner technique (originally developed
in the beginning of the XX century as the so called Janet-Riquier
theory \cite{jan,reid,riq,schwarzJ}) makes this definition
constructive: (\ref{morP}) is a morphism mapping (\ref{sysLPDE}) to
(\ref{sysLPDEQ}) iff for any $i$ the equation $\sum_{j,k}
M_{ij}P_{jk}u_k=0$ is reducible to zero modulo the equations of the
system (\ref{sysLPDE}).

Again, it is easy to see that this category is abelian: for this it
is enough to check that  ${\cal SLPDE}$  is embeddable into the
category of (infinite-dimensional) vector spaces and linear
morphisms and this embedding preserves direct sums, products etc.

It seems natural to refer to Theorem~\ref{JH-abel} to transfer the
many properties of factorization proved for the category ${\cal
LODO}$ in Section~\ref{sec1} to the case of the category  ${\cal
SLPDE}$. Unfortunately this is not so simple: the ascending chains
of monomorphisms are infinite in general: the same example
$|D_x\rangle \subset |D_x,D_y^m\rangle \subset|D_x,D_y^{m-1}\rangle
\subset \ldots |D_x,D_y\rangle \subset |1\rangle$ makes this
obvious. The solution to this problem is given by the standard
construction of a Serre-Grothendieck factorcategory. We refer to
\cite{Faith,Freyd,GM} and especially to \cite{GZ} for a detailed
explanation of this important and general construction. One of the
important steps of this construction is the construction of inverses
of morphisms with ``relatively small'' kernels; the objects are not
formally changed in contrast to the ring-theoretic construction of
factorrings and factormodules. In our case we proceed as follows:
for a given (say, determined) system of LPDE of the form
(\ref{sysLPDE}) (with $s=p$) and take the subcategory ${\cal
S}_{n-2}$ of (overdetermined) systems with solution space
parameterized by functions of at most $n-2$ variables. Then the
Serre-Grothendieck factorcategory ${\cal S}/{\cal S}_{n-2}$ has
finite ascending chains. Another remarkable feature of this
factorcategory is, as we mentioned above, the possibility to
consider morphism which had kernels defined by systems from ${\cal
S}_{n-2}$ as \emph{invertible morphisms}. This may lead to a more
general theory of B\"acklund-type transformations (at least for the
case of linear systems), for example of transformations of Moutard
type (\cite{dar-lec,gour-l}).

Now we can transfer all theoretical results proved in
Section~\ref{sec1} to the case of the factorcategory ${\cal S}/{\cal
S}_{n-2}$ and provide a theoretical foundation for the factorization
theory of arbitrary linear systems of LPDE.

The obvious drawback still lies in the absence of \emph{algorithms}
for such a generalized factorization. We give an overview of
currently known numerous partially algorithmic results in the next
Sections~\ref{ssecLaplace}--\ref{ssecDini}.

%---------------------------------------------------------------------

\subsection{$dim=2$, $ord=2$: Laplace transformations and Darboux integrability
of nonlinear PDEs}\label{ssecLaplace}

Here we expose the basics of the classical theory
\cite{dar-lec,forsyth,gour-l}, which is applicable to hyperbolic
linear partial differential equations of order two with two
independent variables. For simplicity only the case of an equation
with \emph{straight} characteristics will be discussed here:
\begin{equation}\label{Lap}
 Lu =u_{xy} +a(x,y)u_x+b(x,y)u_y +c(x,y)u= 0.
\end{equation}
The more general case can be found in \cite{gour-l,anderson,ts05}.
If one of the \emph{Laplace invariants} of (\ref{Lap})  $ h = a_x +
ab -c$, $k = b_y + ab -c$ vanishes, one can ``naively'' factorize
the operator in the l.h.s.\ of (\ref{Lap}): $k\equiv 0 \Rightarrow
L=\left(D_y +a\right) \left(D_x +b\right)$; $h\equiv 0 \Rightarrow
L=\left(D_x +b\right) \left(D_y +a\right)$.

If $h \neq 0$, $k \neq 0$ then (\ref{Lap}) is \emph{not
factorizable} in the ``naive'' sense. In this case one can perform
one of the two \emph{Laplace transformations} (not to be mixed with
Laplace { transforms}!) which are invertible differential
substitutions (isomorphisms in the category of ${\cal SLPDE}$):
$$ u = \frac{1}{h}   \left(D_x + b\right)u_{(1)}$$
or
$$u = \frac{1}{k}   \left(D_y + a\right)u_{(-1)}.$$
In fact each of the above substitutions is the inverse of the other
up to a functional factor. Each of these substitutions produces a
new operator of the same form (\ref{Lap}) but with different
coefficients and Laplace invariants. The idea of the \emph{Laplace
cascade method} consists in application of these substitutions a few
times, obtaining the (infinite in general) chain
\begin{equation}\label{Lapchain}
      \ldots {\leftarrow} \quad L_{(-2)} \quad {\leftarrow}\quad
    L_{(-1)}\quad {\leftarrow}  \quad   L  \quad {\rightarrow}\quad
     L_{(1)} \quad {\rightarrow} \quad  L_{(2)}\quad  {\rightarrow}
     \ldots
\end{equation}
In some cases (namely these cases are considered as integrable in
this approach) this gives us on some step an operator $L_{(i)}$ with
vanishing $h_{(i)}$ or $k_{(i)}$. Then this chain can not be
continued further in the respective direction and one can find an
explicit formula for the complete solution of the transformed
equation; performing the inverse differential substitutions we
obtain the complete solution of the original equation (with
quadratures).

One of the main results of \cite{ts98} are the following Theorems:
\begin{theorem}
$L=D_x\cdot D_y -a(x,y)D_x -b(x,y)D_y -c(x,y)$ has a nontrivial
generalized right divisor ideal (so is factorizable in the sense
described in Section~\ref{sec2}) iff  the chain (\ref{Lapchain}) of
Laplace transformations is finite at least in one direction.
\end{theorem}
\begin{theorem}
$L=D_x\cdot D_y -a(x,y)D_x -b(x,y)D_y -c(x,y)$ is a lLCM of two
generalized right divisor ideals iff the chain (\ref{Lapchain}) of
Laplace transformations is finite in both directions.
\end{theorem}
This shows the meaning of the generalized definition of \cite{ts98}
and provides a \emph{partial} algorithm for generalized
factorization for equations of the form (\ref{Lap}).

Although practically efficient for simple cases, this method has the
obvious decidability problem: given an operator $L$, how many steps
in the chain (\ref{Lapchain}) should be tried? Currently no stopping
criterion is known. As the example (\ref{cnn+1}) shows, the number
of steps in the chain (equal to $n$ for (\ref{cnn+1}) in the
integrable case $c=n(n+1)$) depends on some subtle arithmetic
properties of the coefficients.

There exists a remarkable link of the theory of Laplace
transformations to the theory of integrable \emph{nonlinear} partial
differential equations. This topic was very popular in the XIX
century and led to the development of integration methods of
Lagrange, Monge, Boole and Ampere. G.~Darboux \cite{darboux}
generalized the method of Monge (known as the method of intermediate
integrals) to obtain the  most  powerful method  for  exact
integration  of partial differential  equations known in the last
century.

Recently in a series of papers \cite{anderson,sokolov,S-Zh}
 the Darboux method was cast into a more precise and efficient (although not
completely algorithmic) form. For the case of a single second-order
nonlinear PDE of the form
\begin{equation}\label{D}
    u_{xy}=F(x,y,u,u_x,u_y)
\end{equation}
the idea consists in linearization: using the substitution $u(x,y)
\rightarrow u(x,y) + \epsilon v(x,y)$ and cancelling terms with
$\epsilon^n$, $n>1$,  we obtain a LPDE
\begin{equation}\label{vxy}
v_{xy} = Av_x + Bv_y + Cv
\end{equation}
with coefficients depending on $x$, $y$, $u$, $u_x$, $u_y$.
Equations of the type (\ref{vxy}) are in fact feasible to the
Laplace cascade method,  certainly one needs to take into
consideration the original equation (\ref{D}) while performing all
the computations of the Laplace invariants and Laplace
transformations: (\ref{D}) allows us to express all the mixed
derivatives of $u$ via $x$, $y$, $u$ and the non-mixed $u_{x\cdots
x}$, $u_{y\cdots y}$). The following statement can be found in
\cite{gour-l}, recently it was rediscovered in
\cite{anderson,sokolov}:
\begin{theorem}  A second order, scalar, hyperbolic
partial differential equation (\ref{D}) is Darboux integrable  if
and only  if  the Laplace  sequence (\ref{Lapchain}) for (\ref{vxy})
is finite in both directions.
\end{theorem}

In \cite{anderson,sokolov} this method was also generalized for the
case of a general second-order nonlinear PDE
$$F(x,y,u,u_x,u_y,u_{xx},u_{xy}, u_{yy})=0
$$.

%---------------------------------------------------------------------

\subsection{$dim=2$, $ord\geq 3$: Generalized Laplace transformations}\label{ssecGenLap}

In \cite{ts05} we have proposed a generalization of the Laplace
cascade method for arbitrary strictly hyperbolic equations with two
independent variables of the form
\begin{equation}\label{3.1}
  \hat L u = \sum_{i+j\leq n} p_{i,j}(x,y)\hat D_x^i\hat D_y^j u =0,
\end{equation}
as well as for $n\times n$ first-order  linear systems
\begin{equation}\label{3.3}
(v_i)_x = \sum_{k=1}^n a_{ik}(x,y)(v_k)_y +
  \sum_{k=1}^n b_{ik}(x,y)v_k
\end{equation}
with strictly hyperbolic matrix $(a_{ik})$.

Here we demonstrate this new method on an example of the
constant-coefficient system
\begin{equation}\label{3x3}
   \left\lbrace  \begin{array}{l}
   D_xu_1= u_1 +2u_2 +u_3, \\
   D_yu_2 = -6u_1 +u_2 +2u_3, \\
   (D_x+D_y)u_3 = 12u_1 +6u_2+u_3.
\end{array}\right.
\end{equation}
It has the following complete explicit solution:
$$ %\! \!\!\!\!\!\!
\left\lbrace  \begin{array}{l}
 u_1 =  %\exp\frac{x+y}{2}\left[
2e^y G(x) + e^{x}(3 F(y) +  F'(y)) +
 \exp\frac{x+y}{2}H(x-y),
%\right], \!\!\!\!\!\!\!\!\!\!\!\!
 \\[0.5em]
 u_2 = e^yG'(x) + 2e^{x} F'(y) - 2u_1,
 \\[0.5em]
 u_3=D_xu_1+3u_1-2(e^yG'(x) + 2e^{x} F'(y)), \\
\end{array}\right.
$$
where $F(y)$, $G(x)$ and $H(x-y)$ are three arbitrary functions of
one variable each.

The solution technology (cf.\ \cite{ts05}) for the details) is again
a differential substitution; in the case of the system (\ref{3x3})
the transformation is given by:
\begin{equation}\label{sol3x3}
 \left\lbrace  \begin{array}{l}
\overline  u_1= u_1, \\
\overline  u_2 =u_2 + 2u_1,\\
\overline  u_3=((D_x+D_y)u_1 -u_1 -2 u_2 - 4u_1 ).
\end{array}\right.
\end{equation}
The transformed system has a triangular matrix and is easily
integrable:
$$   \left\lbrace  \begin{array}{l}
   D_x \overline u_3=  \overline u_3, \\
   D_y\overline u_2 =  2 \overline u_3 + \overline u_2 , \\
   (D_x+D_y)u_1 = \overline u_3 +2 \overline u_2+ u_1.
\end{array}\right.
$$
Again no stopping criterion for the sequences of generalized Laplace
transformations is known in the general case. For constant
coefficient systems an alternative technology was proposed by
F.Schwarz (private communication, 2005): transform the system
(\ref{3x3}) into a Janet (Gr\"obner) normal form with term order:
LEX, $u_3>u_2>u_1,x>y$:
$$
\begin{array}{l}
u_{1,xxy}-u_{1,xx}+u_{1,xyy}-3u_{1,xy}+2u_{1,x}
      -u_{1,yy}+2u_{1,y}-u_1=0,\\
u_{2,y}+3u_2-2u_{1,x}+8u_1=0,\\
u_{2,x}-u_2- \frac{1}{2}u_{1,xx}- \frac{1}{2}u_{1,xy}+3u_{1,x}+
\frac{1}{2}u_{1,y}- \frac{5}{2}u_1=0,\\
 u_3+2u_2-u_{1,x}+u_1=0.
\end{array}
$$
The first equation \emph{factors}:
$$
\begin{array}{l}
D_x^2 D_y - D_x^2 + D_x D_y^2 - 3 D_x D_y + 2 D_x - D_y^2 + 2 D_y
-1\\[1em]
{}\ \ \ \ \ \ \ \ \ \ \  = (D_x + D_y -1)(D_y -1)(D_x -1).
\end{array}
$$
So one can find $u_1$ easily and then the other two functions $u_2$
and $u_3$ are obtained  from the remaining equations of the Janet
base producing essentially the same solution (\ref{sol3x3}).

\textbf{Conjecture}: For  constant-coefficient systems this
Gr\"obner basis technology is equivalent to the generalized Laplace
technology.

%---------------------------------------------------------------------

\subsection{$dim\geq 3$, $ord=2$: Dini transformations}\label{ssecDini}

In \cite{dini1} another simple generalization of Laplace
transformations formally applicable to some second-order operators
in the space of arbitrary dimension was proposed. Namely, suppose
that an operator $\hat L$ has its principal symbol
$$Sym=\sum_{i_1+i_2=2}a_{i_1i_2}(\vec x)D_{x_{i_1}}D_{x_{i_2}}$$
which factors (as a formal polynomial in formal commutative
variables $D_{x_{i}}$) into product of two first-order factors:
$Sym=\hat X_1\hat X_2$ ($\hat X_j=\sum_{i}b_{ij}(\vec x)D_{x_{i}}$
are first-order operators) and moreover the complete operator $\hat
L$ may be written at least in one of the {\em characteristic forms}:
\begin{equation}\label{4}
  \begin{array}{l}
L=  (\hat X_1\hat X_2 + \alpha_1\hat X_1 + \alpha_2\hat X_2 +
  \alpha_3)
  \\
L=  (\hat X_2\hat X_1 + \overline\alpha_1\hat X_1 +
\overline\alpha_2\hat X_2 + \alpha_3),
\end{array}
\end{equation}
where $\alpha_i=\alpha_i(x,y)$. Since the operators $\hat X_i$ do
not necessarily commute we have to take into consideration in
(\ref{4}) and everywhere below the {\em commutation law}
\begin{equation}\label{cl}
  [\hat X_1,\hat X_2] = \hat X_1\hat X_2- \hat X_2\hat X_1 = P(x,y)\hat X_1 +Q(x,y)\hat X_2.
\end{equation}. This is very restrictive
since the two tangent vectors corresponding to the first-order
operators $\hat X_i$ no longer span the complete tangent space at a
generic point $(\vec x_0)$. (\ref{cl}) is also possible only in the
case when these two vectors give an \emph{integrable}
two-dimensional distribution of the tangent subplanes in the sense
of Frobenius, i.e.\ when one can make a change of the independent
variables $(\vec x)$ such that $\hat X_i$ become parallel to the
coordinate plane $(x_1,x_2)$; thus in fact we have an operator $\hat
L$ with only $D_{x_{1}}$,  $D_{x_{2}}$ in it and we have got no
really significant generalization of the Laplace method. If one has
only (\ref{4}) but (\ref{cl}) does not hold one can not perform more
that one step in the Laplace chain (\ref{Lapchain}) and there is no
possibility to get an operator with a zero Laplace invariant (so
naively factorizable and solvable).

Below we demonstrate on an example, following an approach proposed
by U.~Dini in another paper \cite{dini2}, that one can find a better
analogue of Laplace transformations for the case when the dimension
of the underlying space of independent variables is greater than
two. Another particular special transformation was also proposed in
\cite{Athorne}, \cite{YA}; it is applicable to systems whose order
coincides with the number of independent variables. The results of
\cite{Athorne}, \cite{YA} lie beyond the scope of this paper.

Let us take the following equation:
\begin{equation}\label{dex}
 Lu = (D_xD_y + x D_xD_z - D_z)u =0.
\end{equation}
It has three independent derivatives $D_x$, $D_y$, $D_z$, so the
Laplace method is \emph{not} applicable. On the other hand its
principal symbol splits into product of two first-order factors:
$\xi_1\xi_2 + x \xi_1\xi_3 =\xi_1(\xi_2+x\xi_3)$. This is no longer
a typical case for hyperbolic operators in dimension~$3$; we will
use this special feature introducing two characteristic operators
$\hat X_1=D_x$, $\hat X_2=D_y + x D_z$. We have again a nontrivial
commutator  $[\hat X_1,\hat X_2] =  D_z= \hat X_3$. The three
operators $\hat X_i$ span the complete tangent space in every point
$(x,y,z)$. Using them one can represent the original second-order
operator in one of two partially factorized forms:
$$ L = \hat X_2\hat X_1 - \hat X_3 =  \hat X_1\hat X_2 - 2\hat X_3.$$
Let us use the first one and transform the equation into a system of
two first-order equations:
%$$
%Lu=0 \Longleftrightarrow
%(D_y + xD_z)\underbrace{D_x u}_v - D_z u =0,
%$$
\begin{equation}\label{Dini2e}
 Lu=0 \Longleftrightarrow
   \left\lbrace  \begin{array}{l}
\hat X_1 u = v, \\
\hat X_3 u = \hat X_2 v.
\end{array}\right.
\end{equation}
Cross-differentiating the left hand sides of (\ref{Dini2e}) and
using the obvious identity $[\hat X_1,\hat X_3] = [ D_x, D_z]=0$ we
get $  \hat X_1 \hat X_2v =   D_x (D_y + xD_z) v = \hat X_3 v=D_z v
$ or $ 0=D_x (D_y + xD_z) v - D_z v = (D_x D_y + x D_x D_z) v
 = (D_y + xD_z)  D_x v = \hat X_2\hat X_1 v$.

This is precisely the procedure proposed by Dini in \cite{dini2}.
Since it results now in another second-order equation which is
``naively'' factorizable we easily find its complete solution:
$$v= \int \phi(x,xy-z) \, dx + \psi(y,z)$$
where $\phi$ and $\psi$ are two arbitrary functions of two variables
each; they give the general solutions of the equations $\hat
X_2\phi=0$, $\hat X_1\psi=0$.

Now we can find  $u$:
$$ u= \int \Big(v\, dx  + (D_y + xD_z)v\, dz \Big)+ \theta(y),
$$
where an extra free function $\theta$ of one variable appears as a
result of integration in (\ref{Dini2e}).

So we have seen that such {\em Dini transformations} (\ref{Dini2e})
in some cases may produce a complete solution in explicit form for a
non-trivial three-dimensional equation (\ref{dex}). This explicit
solution can be used to solve initial value problems for
(\ref{dex}).

Dini did not give any general statement on the range of
applicability of his trick. In \cite{ts06} we have proved the
following
\begin{theorem}
 Let $L=\sum_{i+j+k\leq 2}a_{ijk}(x,y,z)D_x^iD_y^jD_z^k$ have
factorizable principal symbol: $\sum_{i+j+k=
2}a_{ijk}(x,y,z)D_x^iD_y^jD_z^k= \S_1\S_2$ (mod lower-order terms)
with generic (non-commuting) first-order LPDO $\S_1$, $\S_2$. Then
there exist two Dini transformations $L_{(1)}$, $L_{(-1)}$ of $L$.
\end{theorem}
\textsl{Proof.} One can represent $L$ in two possible ways:
\begin{equation}\label{th-1}
L=\S_1\S_2 + \T + a(x,y,z) =\S_2\S_1 + \hat U + a(x,y,z)
\end{equation}
with some first-order operators $\T$, $\hat U$. We will consider the
first one obtaining a transformation of $L$ into an operator
$L_{(1)}$ of similar form.

In the  generic case the operators $\S_1$, $\S_2$, $\T$ span the
complete 3-dimensional tangent space in a generic point $(x,y,z)$.
Precisely this requirement will be assumed to hold hereafter;
operators $L$ with this property will be called {\em generic}.

Let us fix the coefficients in the expansions of the following
commutators:
\begin{equation}\label{st-comm1}
    [\S_2, \T] = K(x,y,z)\S_1 +M(x,y,z)\S_2 + N(x,y,z)\T.
\end{equation}
\begin{equation}\label{st-comm2}
    [\S_1, \S_2] = P(x,y,z)\S_1 +Q(x,y,z)\S_2 + R(x,y,z)\T.
\end{equation}

First we try to represent the operator in a  partially factorized
form: $L=(\S_1 + \a)(\S_2 +\b)  + \V + b(x,y,z)$ with some
indefinite $\a=\a(x,y,z)$, $\b=\b(x,y,z)$ and $\V=\T-\b\S_1
-\a\S_2$, $b=a-\a\b-\S_1(\b)$.

Then introducing $v=(\S_2 +\b)u$ we get the corresponding
first-order system:
\begin{equation}\label{Dini2}
Lu=0 \Longleftrightarrow \left\lbrace  \begin{array}{l}
(\S_2 +\b)u = v, \\
(\V+b) u = - (\S_1 + \a)v.
\end{array}\right.
\end{equation}
Next we try to eliminate $u$ by cross-differentiating the left hand
sides, which gives
\begin{equation}\label{th-L1}
[(\V+b), (\S_2 +\b) ]u = (\S_2 +\b)(\S_1 + \a)v +(\V+b)v.
\end{equation}
If one wants $u$ to disappear from this new equation one should find
out when $[(\V+b), (\S_2 +\b)]u$ can be transformed into an
expression involving \emph{only} $v$, i.e.\ when this commutator is
a linear combination of just two expressions $(\S_2 +\b)$ and
$(\V+b)$:
\begin{equation}\label{th-comm}
    [(\V+b), (\S_2 +\b)] = \mu(x,y,z)(\S_2 +\b) + \nu(x,y,z)(\V+b).
\end{equation}
This is possible to achieve choosing the free functions $\a(x,y,z)$,
$\b(x,y,z)$ appropriately. In fact, expanding the left and right
hand sides in (\ref{th-comm}) in the local basis of the initial
fixed operators $\S_1$, $\S_2$, $\T$ and the zeroth-order operator
$1$ and collecting the coefficients of this expansion, one gets the
following system for the unknown functions $\a$, $\b$, $\mu$, $\nu$:
%\begin{equation}\label{sys-exp}
$$
 \left\lbrace  \begin{array}{l}
    K  +\b P - \S_2(\b) = \nu \b   ,     \\
    M  -\S_2(\a) +\b Q  =   \nu \a - \mu ,\\
    N + \b R = - \nu       ,  \\
    \b\S_1(\b)-\T(\b)+\S_2(a) -\b\S_2(\a) -\S_2(\S_1(\b)) =
          - \nu (a-\a\b-\S_1(\b))-\mu\b.
 \end{array}\right.
$$
%\end{equation}
After elimination of $\nu$ from its first and third equations we get
a first-order non-linear partial differential equation for $\b$:
\begin{equation}\label{eq-b}
 \S_2(\b) = \b^2R+(N+P)\b +K.
\end{equation}
This Riccati-like equation may be transformed into a second-order
linear PDE via the standard substitution $\b = \S_2(\gamma)/\gamma$.
Taking any non-zero solution $\b$ of this equation and substituting
$\mu = \nu \a +\S_2(\a) -\b Q -M $ (taken from the second equation
of the system) into the fourth equation of the system we obtain a
first-order linear partial differential equation for $\a$ with the
first-order term $\b\S_2(\a)$. Any solution of this equation will
give the necessary value of $\a$. Now we can substitute $[(\V+b),
(\S_2 +\b)]u = \mu(\S_2 +\b)u + \nu(\V+b)u= \mu v -\nu (\S_1 + \a)v$
into the left hand side of (\ref{th-L1}) obtaining the transformed
equation $L_{(1)}v=0$.

If we would start the same procedure using the second partial
factorization in (\ref{th-1}) we would find the other transformed
equation  $L_{(-1)}w=0$.
 $\Box$

%---------------------------------------------------------------------

\section{Other results and conjectures}

The theory of integration of linear and nonlinear partial
differential equations was among the most popular topics in the XIX
century. Enormous amount of papers were devoted for example to
transformations of equations to an integrable form. In particular
the papers \cite{LeRoux,pisati,Petren} were devoted to a more
general Laplace type transformations. Some of these results were
obtained in the framework of the classical differential geometry;
cf.\ \cite{fer-lap,K-T1} for a modern exposition of those results.

In addition to the problems studied above one should mention a class
of overdetermined systems of linear partial differential equations
\textit{with finite-dimensional solution space} studied in
\cite{ZSchTs03,minwu}. There an algorithm for factorization of such
systems was proposed.

Another popular in the past decade topic was the theory of ``naive''
factorization, i.e.\ representation of a given LPDO as a product of
lower-order LPDO: in \cite{GSch04} an algorithm for such
factorization was proposed for the case of operators with symbol
representable as a product of two coprime polynomials. This result
was developed further in \cite{SW}.

From the theory of Laplace and Dini transformations the following
\textbf{conjectures} seem to be natural:

\begin{itemize}
    \item \emph{If a LPDO is factorizable in the generalized sense, then its
principal symbol is factorizable as a multivariate commutative
polynomial.}

\item
\emph{If a LPDO of order $n$ has a complete solution in a
quadrature-free form (\ref{complSol}) then its symbol splits into
$n$ linear factors.}

\end{itemize}

\section{Acknowledgment}

The author enjoys the occasion to thank the organizers of the LMS
summer lecture course and the complete mini-program ``Algebraic
Theory of Differential Equations'' at the International Centre for
Mathematical Sciences, Edinburgh for their efforts which guaranteed
the success of the mini-program as well as for partial financial
support which made presentation of the results given above possible.

\end{document}